\documentstyle[amstex,preprint,aps,epsfig]{revtex}
\begin{document}

%la instruccion preprint debe venir justo destras de esta linea.

\preprint{FTUV-00-1027}

\title{The $\beta \beta 2\nu $-decay in $^{48}Ca$}
\author{M. T. Capilla$^{1}\thanks{%
Email: capilla at titan.ific.uv.es}$ , B. Desplanques$^{2}\thanks{%
Email: desplanq at isn.in2p3.fr}\ $ and S. Noguera$^{1}\thanks{%
Email: Santiago.Noguera at uv.es}$}
\address{1)Departamento de F\'{i}sica Te\'{o}rica; Universidad de Valencia;\\
c/ Dr. Moliner 50; E-46100 Burjassot(Valencia); Spain\\
2)Institut des Sciences Nucl\'{e}aires, UMR CNRS-UJF,\\
F-38026 Grenoble Cedex, France}
\maketitle

\begin{abstract}
A schematic study of the $\beta \beta 2\nu $-decay of $^{48}Ca$ is made in a
shell-model approach. The emphasis is especially put on the role of the
spin-orbit potential in relation with the contribution of other terms in the
strong interaction. This is discussed with a particular attention to the
behavior of these ones under the $SU(4)$ symmetry. Different methods in
calculating the transition amplitude are also looked at with the aim to
determine their reliability and, eventually, why they don't work. Further
aspects relative to the failure of the Operator Expansion Method to
reproduce the results of more elaborate calculations are examined.
\end{abstract}

\section{INTRODUCTION}

The importance of the double beta decay process is well recognized. First,
the neutrinoless mode, yet unobserved, is of fundamental interest, as it
will be a signal for neutrino mass and lepton number non-conservation.
Second, the double beta decay with two-neutrino emission ($\beta \beta 2\nu
),$ allowed in the standard model, is a very rare process which has not been
experimentally observed until 1987, in the decay of $^{82}Se$ studied by
Elliot et al. \cite{Elliot}. Subsequent results in other nuclei were
obtained by other groups (see \cite{Morales} for a recent review of the
experimental situation). Recently Balysh et al. \cite{Balysh} have measured
the double beta decay half-life of $^{48}Ca.$ This nucleus is the lightest
one for which such a measurement is feasible.

On the theoretical side (see \cite{Suhonen,Faessler} for a recent review)
the $\beta \beta 2\nu $-decay which, at the beginning, was a well defined
process in the standard model, has revealed as a real challenge for nuclear
model practitioners. There are two reasons to this situation. On the one
hand, the decay mode is highly suppressed and sensitively depends on poorly
determined parts of the nuclear interaction. On the other, it is a second
order process, which implies a summation on intermediate, and not always
well determined, states. Thus, even if it is not a process involving new
fundamental physics, the $\beta \beta 2\nu $-decay is related to a new type
of nuclear matrix element. This one incorporates information on the wave
functions that is not given by other standard observables.

The difficulties in the calculation of the $\beta \beta 2\nu $-decay have
been expressed in several different but related ways in the literature:

\begin{itemize}
\item  In QRPA calculations, it has been related to the extreme sensitivity
of the transition matrix element to the so-called $g_{pp}$ parameter, which
governs the $pn$-excitations \cite{Vogel,Engel,Krmpotic}.

\item  In the $SU(4)$ language, it has been connected to poor determination
of the nuclear force in the $L=0,\,S=1,\,T=0$ ($J=1^{+}\,T=0)$\ channel \cite
{Bernabeu,Krmpotic}. At the same time, it was also observed in this scheme
that strong truncation of the model basis can produce undesirable
contributions to the transition matrix element.

\item  It has been also connected to the bad description of the $\beta ^{+}$
decays, which processes have been used to fit the unknown parts of the
nuclear force \cite{Engel}.
\end{itemize}

In order to avoid some of the previous uncertainties, an alternative
approach has been proposed, the Operator Expansion Method (OEM) \cite
{Simko,Ching,Ching-Klapdor}.

In the present paper, we will focus our attention on the role of different
parts of the nuclear force in the $\beta \beta 2\nu $-decay as well as on
different methods used in the literature to describe this process. For that
purpose, the simplest nuclear transition to be studied is the $\beta \beta
2\nu $-decay in $^{48}Ca$, that offers a double advantage. There exist both
a sensitive experimental value $\left( T_{1/2}^{2\nu }=\left( 4.3%
%TCIMACRO{\QATOPD. . {+2.4}{-1.1}}%
%BeginExpansion
{+2.4 \atopwithdelims.. -1.1}%
%EndExpansion
\left[ Stat\right] \pm 1.4\left[ Syst\right] \right) \times 10^{19}%
%TCIMACRO{\unit{y}}%
%BeginExpansion
\mathop{\rm y}%
%EndExpansion
\cite{Balysh}\right) $ and an elaborate shell model calculation \cite
{Caurier2}, which is the natural calculation scheme for this nucleus. While
doing these studies, we will have in mind a long standing problem. Different
calculations were approximately leading to the same decay rate whereas the
intrinsic sign of the transition matrix element was not the same \cite
{Bernabeu2}, requiring some clarification. With this respect, we will in
particular show how higher order effects in the $SU(4)$ symmetry breaking
interaction modify previous estimates. On the other hand, a critical study
of the OEM approach has been done in \cite{Engel1}. Based on an analysis of
the $SU(4)$ symmetry breaking effects, other features of the OEM have been
revealed, which deserve discussion. Our work is therefore concerned more
with the role of various approximations than with a realistic calculation of
the process. We will use an analytical force to achieve this objective. This
allows us to easily compare various methods of calculation and to switch on
and off the different parts of the interaction.

The plan of the paper is as follows. We remind in the second section
expressions for the standard transition operator and that one obtained in
the OEM approach. In section 3, we introduce in the OEM expression the
Coulomb splitting effect and make a comparison of the corresponding result
with the transition operator derived independently at the first order in the 
$SU(4)$ symmetry breaking interaction. The effective NN interaction that we
will use for our study is specified in section 4. The fifth section is
devoted to a presentation of our results together with a discussion.

\section{THE $\protect\beta\protect\beta2\protect\nu$-DECAY: DESCRIPTION OF
THE TRANSITION OPERATOR}

The $\beta \beta 2\nu $-decay is allowed in the Standard Model. In this
process, two neutrons decay to two protons with the emission of two
electrons and two neutrinos. The Lagrangian responsible for that process is
the standard Fermi one. Under the usual assumptions (the impulse
approximation is assumed, the lepton energies are replaced by their average
values and non-conservation of isospin is discarded), we obtain that the
life time is expressed by 
\begin{equation}
\left[ T_{1/2}^{\beta \beta 2\nu }\right] ^{-1}=G_{2\nu }\left| M_{{\cal GT}%
}\right| ^{2},
\end{equation}
where $G_{2\nu },$ which can be found in \cite{Suhonen,Doi}, contains all
the leptonic part and the integral on the phase space. The nuclear
information is included in the nuclear matrix element 
\begin{equation}
M_{{\cal GT}}=\frac{i}{2}\int_{0}^{\infty }dt\,e^{i\Delta t}<0_{f}^{+}|\left[
e^{iHt}\vec{A}e^{-iHt},\vec{A}\right] |0_{i}^{+}>,  \label{MGT}
\end{equation}
with $\vec{A}=\sum \vec{\sigma}_{i}\tau _{i}^{+},$ the Gamow-Teller
operator, and $\Delta =\frac{1}{2}\left( E_{i}-E_{f}\right) $.

Methods for evaluating the $\beta \beta 2\nu $-decays differ in the
approximations made in order to calculate (\ref{MGT}). What we will call the
standard method consists to insert a complete set of intermediate states in
the commutator (\ref{MGT}). This allows one to perform the time integral and
we obtain 
\begin{equation}
M_{{\cal GT}}^{St}=\sum_{n}\frac{<0_{f}^{+}|\vec{A}|1_{n}^{+}><1_{n}^{+}|%
\vec{A}|0_{i}^{+}>}{E_{n}-\frac{1}{2}\left( E_{i}+E_{f}\right) }.
\label{MGTst}
\end{equation}

Theoretically this method is exact, but in practical calculations some
limitation occurs. Estimating the matrix element given by Eq. (\ref{MGTst})
implies to consider all the intermediate states contributing to the sum.
Thus, the method could not be the most interesting one if the intermediate
states or their energies cannot be well determined and if cancellations
between different contributions are present. As we will explain later, the
approximated $SU\left( 4\right) $ symmetry of the nuclear forces tells us
that such cancellations must be present in (\ref{MGTst}), leading to
uncertainties in employing this method of calculation.

On the other hand, in general, nuclei which undergo double beta decay are
open shell nuclei and the usual formalism for describing them is the QRPA.
In that case, there are new difficulties in the evaluation of the $M_{{\cal %
GT}}^{St}$ amplitude due to the fact that the QRPA is near the collapse for
the physical values of the so-called $g_{pp}$ parameter. Improvements of the
QRPA, like full-QRPA, renormalized QRPA or full-RQRPA, can solve these
problems but other ones arrive, like the violation of the Ikeda sum rule
(see discussion in ref. \cite{Suhonen,Faessler}, and see also \cite{Hirsch}).

As mentioned in the introduction, the above difficulties have been a source
of concern. An alternative approach to calculate directly the exponentials
appearing in (\ref{MGT}), the operator expansion method (OEM) \cite
{Simko,Ching,Ching-Klapdor}, has thus been proposed some years ago. This is
impossible in a general manner and the main assumption made by the authors
is to only retain two-body operators, which supposes that only the two
nucleons involved in the transition are of special interest in the
calculation. This is like a spectator approximation in the sense that all
the nucleons not involved in the $\beta \beta $ transition don't contribute
to the process. On the other hand, the interaction between the two active
nucleons is included to all orders, neglecting some parts as we will explain
later. A diagrammatic view of this approach is shown in Fig. \ref{fig1}a.

In the simplest approximation, the kinetic energy term, the spin-orbit and
the tensor potentials are neglected in the Hamiltonian $H$ appearing in (\ref
{MGT}). Under these assumptions, only the central part of the potential
contributes. Starting from its expression written as follows 
\begin{equation}
H\approx V_{c}=\sum_{i>j}\left[ v_{o}\left( r\right) +v_{\tau }\left(
r\right) \vec{\tau}_{i}.\vec{\tau}_{j}+v_{\sigma }\left( r\right) \vec{\sigma%
}_{i}.\vec{\sigma}_{j}+v_{\sigma \tau }\left( r\right) \vec{\tau}_{i}.\vec{%
\tau}_{j}\,\vec{\sigma}_{i}.\vec{\sigma}_{j}\right] \,\,,
\label{central pot}
\end{equation}
the $\beta \beta 2\nu $-transition amplitude in the OEM approach, first
derived by \v{S}imkovic\ et al. \cite{Simko} and Ching et al.\ \cite{Ching},
can be expressed as 
\begin{equation}
M_{{\cal GT}}^{OEM}=<0_{f}^{+}|\sum_{i>j}{\cal M}_{i,j}\tau _{i}^{+}\tau
_{j}^{+}|0_{i}^{+}>,  \label{MGToem}
\end{equation}
with 
\begin{eqnarray}
{\cal M}_{ij} &=&\frac{24\left[ v_{\sigma }\left( r\right) -v_{\tau }\left(
r\right) \right] }{\Delta ^{2}-16\left[ v_{\sigma }\left( r\right) -v_{\tau
}\left( r\right) \right] ^{2}}\Omega _{0}\left( ij\right)  \nonumber \\
&&+\frac{8\left[ 2v_{\sigma \tau }\left( r\right) -v_{\sigma }\left(
r\right) -v_{\tau }\left( r\right) \right] }{\Delta ^{2}-16\left[ 2v_{\sigma
\tau }\left( r\right) -v_{\sigma }\left( r\right) -v_{\tau }\left( r\right) %
\right] ^{2}}\Omega _{1}\left( ij\right) ,  \label{MGToem1}
\end{eqnarray}
where $\Omega _{0,1}$ represents the projector operator on spin 0 and 1
subspaces.

The Gamow-Teller operator is a $SU(4)$ generator. Then, from equation (\ref
{MGT}), we observe that $M_{{\cal GT}}$ will be zero if the Hamiltonian were 
$SU(4)$ invariant (we discard transitions between members of the same $%
SU\left( 4\right) $ multiplets). This result is not obvious from (\ref{MGTst}%
) where, in the general case, each intermediate state could give a non-zero
contribution, the zero being obtained after summation over all intermediate
states. Usual nuclear potentials are not so far from the $SU(4)$ symmetry,
then we must expect cancellations in the summation present in (\ref{MGTst}).
From this point of view, expression (\ref{MGToem1}) is more transparent,
because the condition over the central force to be $SU(4)$ symmetric is just 
$v_{\sigma }\left( r\right) =v_{\tau }\left( r\right) =v_{\sigma \tau
}\left( r\right) $. Notice that in a particular case, actually close to most
realistic transitions, the vanishing of the matrix element would result from
the fact that the operator $\vec{A}$ in Eq. (\ref{MGTst}) acting on the
final state gives zero.

\section{OEM AND SU(4) APPROXIMATION}

We can improve the OEM model, Eq. (\ref{MGToem1}), by incorporating the
contribution of the Coulomb interaction and then make a comparison of the
expression so obtained with that one derived in the first order $SU(4)$
symmetry breaking approximation.

The Coulomb interaction is an important ingredient of the spectroscopy of
medium-heavy nuclei involving different charges as it provides a few MeV
shifts, which compare to the energy splittings produced by the strong
interaction itself. For our purpose, we added to this interaction, $H_{s}$,
a constant term proportional to the third component of the total isospin of
the nucleus 
\begin{equation}
H=H_{s}+\Delta_{c}T_{3}.  \label{Coulomb Force}
\end{equation}

The above Coulomb force can be easily included in the operator appearing in (%
\ref{MGToem}). Using that $\left[ T_{3},\vec{A}\right] =\vec{A}$, we obtain
the modification of (\ref{MGToem1}) into 
\begin{align}
{\cal M}_{ij}& =\frac{24\left[ v_{\sigma }\left( r\right) -v_{\tau }\left(
r\right) \right] }{\left( \Delta +\Delta _{c}\right) ^{2}-16\left[ v_{\sigma
}\left( r\right) -v_{\tau }\left( r\right) \right] ^{2}}\Omega _{0}\left(
ij\right)  \nonumber \\
& +\frac{8\left[ 2v_{\sigma \tau }\left( r\right) -v_{\sigma }\left(
r\right) -v_{\tau }\left( r\right) \right] }{\left( \Delta +\Delta
_{c}\right) ^{2}-16\left[ 2v_{\sigma \tau }\left( r\right) -v_{\sigma
}\left( r\right) -v_{\tau }\left( r\right) \right] ^{2}}\Omega _{1}\left(
ij\right) .  \label{MGToem2}
\end{align}
From the definition of $\Delta $ and (\ref{MGToem1}), we have the relation, $%
\Delta +\Delta _{c}=\frac{1}{2}\left( E_{i}^{s}-E_{f}^{s}\right) ,$ where $%
E_{i}^{s}$ is the strong interaction contribution to the energy of the state.

As we said before, the nuclear forces are not far from the $SU(4)$ symmetry
and, if that symmetry were exact, the double beta transition amplitude will
be zero when it connects states belonging to different $SU(4)$ multiplets.
In order to look at the consistency of our results, it can be useful to
study the first order correction of our expressions in the $SU(4)$ breaking
parts of the force. Let us write for that 
\begin{equation}
H=H_{0}+H_{1},
\end{equation}
where $H_{0}$ $(H_{1})$ represents the $SU(4)$ symmetric (breaking) part of
the force. The hamiltonian $H_{0}$ is a purely central force and has two
terms in the spin-isospin space, one proportional to $1$ and the other
proportional to the Casimir of the $SU(4)$ group, $\left( \vec{\tau}_{i}.%
\vec{\tau}_{j}+\vec{\sigma}_{i}.\vec{\sigma}_{j}+\vec{\tau}_{i}.\vec{\tau}%
_{j}\,\vec{\sigma}_{i}.\vec{\sigma}_{j}\right) $%
\begin{equation}
H_{0}=\sum_{i>j}\left[ v_{o}\left( r\right) +\frac{1}{5}\left( v_{\tau
}\left( r\right) +v_{\sigma }\left( r\right) +3v_{\sigma \tau }\left(
r\right) \right) \left( \vec{\tau}_{i}.\vec{\tau}_{j}+\vec{\sigma}_{i}.\vec{%
\sigma}_{j}+\vec{\tau}_{i}.\vec{\tau}_{j}\,\vec{\sigma}_{i}.\vec{\sigma}%
_{j}\right) \right] .
\end{equation}
The Hamiltonian, $H_{1}$, which is able to give a double beta transition at
the first order in the $SU(4)$ symmetry, involves two other combinations of
the components appearing in the central force, Eq. (\ref{central pot}), 
\begin{align}
H_{1}& =\sum_{i>j}\left[ \frac{1}{2}\left( v_{\tau }\left( r\right)
-v_{\sigma }\left( r\right) \right) \left( \vec{\tau}_{i}.\vec{\tau}_{j}-%
\vec{\sigma}_{i}.\vec{\sigma}_{j}\right) \right.  \nonumber \\
& +\left. \frac{3}{10}\left( v_{\tau }\left( r\right) +v_{\sigma }\left(
r\right) -2v_{\sigma \tau }\left( r\right) \right) \left( \vec{\tau}_{i}.%
\vec{\tau}_{j}+\vec{\sigma}_{i}.\vec{\sigma}_{j}-\frac{2}{3}\vec{\tau}_{i}.%
\vec{\tau}_{j}\,\vec{\sigma}_{i}.\vec{\sigma}_{j}\right) \right] .
\label{H1}
\end{align}
Starting with (\ref{MGT}), we observe that there are two different
situations. First, let us consider a nucleus with only two active nucleons.
In that case it is obvious that $\left[ H_{0},H_{1}\right] =0,$ and the
exponential of the Hamiltonian present in (\ref{MGT}) can be splitted in two
exponentials relative to $H_{0}$ and $H_{1}$ respectively. The first one,
related to $H_{0},$ commutes with all the operators and can be ruled out.
The second one, related to $H_{1},$ can be expanded to the first order in $%
H_{1}$ and we obtain 
\begin{equation}
M_{{\cal GT}}^{(2neutrons)}=\frac{1}{2\left( \Delta +\Delta _{c}\right) ^{2}}%
<0_{f}^{+}|\left[ \left[ H_{1},\vec{A}\right] ,\vec{A}\right] |0_{i}^{+}>.
\label{MGToem.su4}
\end{equation}
Calculating explicitly these commutators, we get 
\[
M_{{\cal GT}}^{(2neutrons)}=\frac{1}{\left( \Delta +\Delta _{c}\right) ^{2}}%
<0_{f}^{+}|\sum_{i>j}\left\{ 24\left[ v_{\sigma }\left( r\right) -v_{\tau
}\left( r\right) \right] \Omega _{0}\left( ij\right) \right. 
\]
\begin{equation}
+\left. 8\left[ 2v_{\sigma \tau }\left( r\right) -v_{\sigma }\left( r\right)
-v_{\tau }\left( r\right) \right] \Omega _{1}\left( ij\right) \right\} \tau
_{i}^{+}\tau _{j}^{+}|0_{i}^{+}>.
\end{equation}
This result agrees with the Coulomb corrected OEM expression, Eq. (\ref
{MGToem2}), when this one is expanded up to the first order in the $SU(4)$
symmetry breaking.

Surprisingly, a different result is obtained when we consider a nucleus with
more than two valence nucleons. In view of its importance for a comparison
with the above OEM result, Eq. (\ref{MGToem2}), we give here some detail on
its derivation.

Beyond the two valence nucleon case, $\left[ H_{0},H_{1}\right] $ doesn't
vanish and from (\ref{MGT}) it can be shown that, up to first order in $%
H_{1} $, 
\begin{equation}
M_{{\cal GT}}^{SU(4)}=\frac{i}{2}\int_{0}^{\infty }dt\,e^{i\left( \Delta
+\Delta _{c}\right) t}\sum_{n=1}^{\infty }\frac{\left( it\right) ^{n}}{n!}%
<0_{f}^{+}|%
%TCIMACRO{\underset{n-1}{\underbrace{[H_{0,}...[H_{0},}}}%
%BeginExpansion
\mathrel{\mathop{\underbrace{[H_{0,}...[H_{0},}}\limits_{n-1}}%
%EndExpansion
[[H_{1},\vec{A}],\vec{A}]]...]|0_{i}^{+}>.  \label{MGTsu4.0}
\end{equation}
In order to perform the sum present in Eq. (\ref{MGTsu4.0}), let us define
an operator ${\cal B=}\left[ \left[ H_{1},\vec{A}\right] ,\vec{A}\right] $
and introduce an operator ${\cal C}$ solution of the equation ${\cal B}=%
\left[ H_{0},{\cal C}\right] .$ In terms of the ${\cal C}$ operator, (\ref
{MGTsu4.0}) can be rewritten as 
\begin{eqnarray}
M_{{\cal GT}}^{SU(4)} &=&\frac{i}{2}\int_{0}^{\infty }dt\,e^{i\left( \Delta
+\Delta _{c}\right) t}\sum_{n=1}^{\infty }\frac{\left( it\right) ^{n}}{n!}%
<0_{f}^{+}|%
%TCIMACRO{\underset{n}{\underbrace{[H_{0,}...[H_{0},}}}%
%BeginExpansion
\mathrel{\mathop{\underbrace{[H_{0,}...[H_{0},}}\limits_{n}}%
%EndExpansion
{\cal C}]...]|0_{i}^{+}>  \nonumber \\
&=&\frac{i}{2}\int_{0}^{\infty }dt\,e^{i\left( \Delta +\Delta _{c}\right)
t}<0_{f}^{+}|\left[ e^{iH_{0}t}{\cal C}e^{-iH_{0}t}{\cal -C}\right]
|0_{i}^{+}> \\
&=&\frac{1}{\Delta +\Delta _{c}}<0_{f}^{+}|{\cal C}|0_{i}^{+}>.
\end{eqnarray}
The solution of the equation ${\cal B}=\left[ H_{0},{\cal C}\right] $ is 
\begin{equation}
{\cal C}=-i\lim_{\varepsilon \rightarrow 0}\int_{0}^{\infty }dq\,e^{iH_{0}q}%
{\cal B\,}e^{-iH_{0}q}\,e^{-\varepsilon q}.
\end{equation}
Performing this last integration and using the explicit expression of ${\cal %
B}$, we obtain: 
\begin{equation}
M_{{\cal GT}}^{SU(4)}=-\frac{1}{2\left( \Delta +\Delta _{c}\right) ^{2}}%
<0_{f}^{+}|\left[ \left[ H_{1},\vec{A}\right] ,\vec{A}\right] |0_{i}^{+}>.
\label{MGTsu4}
\end{equation}
This equation can also be obtained from (\ref{MGTst}). To do that, we must
realize that up to first order in the $H_{1}$ Hamiltonian, the double beta
decay implies a transition between different $SU(4)$ multiplet states. In
the case of interest here, it involves the state $\left| \left[ 4,4\right]
T=4,S=0\right\rangle $ associated with the $^{48}Ca$ ground state and the
state $\left| \left[ 2,2\right] T=2,S=0\right\rangle $ related to the $%
^{48}Ti$ \cite{Bernabeu}.\ Then, there is only one intermediate state
contributing to the amplitude, which is the Gamow-Teller resonance of the
initial state (with an energy $E_{n}=E_{i}+\Delta _{c})$: 
\begin{equation}
M_{{\cal GT}}^{SU(4)}=\sum_{n}\frac{<0_{f}^{+}|\vec{A}|1_{n}^{+}><1_{n}^{+}|%
\vec{A}|0_{i}^{+}>}{\left( E_{i}+\Delta _{c}\right) -\frac{1}{2}\left(
E_{i}+E_{f}\right) }=\frac{1}{\left( \Delta +\Delta _{c}\right) }<0_{f}^{+}|%
\vec{A}.\vec{A}|0_{i}^{+}>,  \label{su4.4}
\end{equation}
and $H_{1}$ appears in the mixing in the final nucleus between the two $%
SU(4) $ representations 
\begin{equation}
|0_{f}^{+}>=|0_{f}^{+}>_{0}+\sum_{r}\frac{1}{\left( E_{f}-E_{r}\right) }%
|0_{r}^{+}>_{0}\,_{0}\hspace{-0.15cm}<0_{r}^{+}|H_{1}|0_{f}^{+}>_{0},
\label{su4.5}
\end{equation}
where $|0_{f}^{+}>_{0}$ is the pure $SU\left( 4\right) $ final state. When
we introduce (\ref{su4.5}) in (\ref{su4.4}), only states $|0_{r}^{+}>_{0}$
belonging to the same $SU(4)$ supermultiplet as $%
|0_{i}^{+}>(=|0_{i}^{+}>_{0})$ can give non-zero contribution and the
energies of these states are $E_{r}=E_{i}+2\Delta _{c}.$ Then we obtain 
\begin{align}
M_{{\cal GT}}^{SU(4)}& =\frac{1}{\left( \Delta +\Delta _{c}\right) }\sum_{r}%
\frac{1}{\left( E_{f}-E_{r}\right) }\,\,\,_{0}\hspace{-0.15cm}%
<0_{f}^{+}|H_{1}|0_{r}^{+}>_{0}\,_{0}\hspace{-0.15cm}<0_{r}^{+}|\vec{A}.\vec{%
A}|0_{i}^{+}>_{0}  \nonumber \\
& =-\frac{1}{2\,\left( \Delta +\Delta _{c}\right) ^{2}}\,\,_{0}\hspace{%
-0.15cm}<0_{f}^{+}|H_{1}\vec{A}.\vec{A}|0_{i}^{+}>_{0}.
\end{align}
This result is in agreement with Eq. (\ref{MGTsu4}) because the other terms
present in the double commutator, $<0_{f}^{+}|\left[ \left[ H_{1},\vec{A}%
\right] ,\vec{A}\right] |0_{i}^{+}>$, vanish in the $SU(4)$ limit. The main
point here is that this expression has a sign opposite to (\ref{MGToem.su4}%
). This is due to\ the presence of many nucleons operators in the former
expression, as we represent in Fig. \ref{fig1}b, while the latter one only
contains two-body operators with the consequence to provide the wrong sign
in the first order $SU(4)$ symmetry breaking limit. In particular,
contributions due to pure Pauli antisymmetrization, as those depicted in
Fig. \ref{fig1}c, are not accounted for in the OEM.

More important, in the simplest case where the operator $H_{0}$ in Eq. (\ref
{MGTsu4.0}) can be approximated by the sum of the single particle energies,
the different commutators appearing in this expression can be calculated.
Their contributions, which form a non-convergent geometrical series, are
given, up to a factor, by the sum

\begin{equation}
\frac{1}{2}(1+2+4+8+...)(...),
\end{equation}
where the first term in the parentheses is that one retained by the OEM. To
get these contributions, we used the relation, $E_{i}^{s}-E_{f}^{s}=2\left(
\Delta +\Delta _{c}\right) $. Formally, the above sum can be performed with
the result 
\begin{equation}
\frac{1}{2}\,\frac{1}{1-2}(...)=-\frac{1}{2}(...).
\end{equation}
This is the result obtained from a direct calculation, Eq. (\ref{MGTsu4}).
It specifies in two ways the failure of the OEM demonstrated on a
quantitative basis by Engel et al. \cite{Engel1}. i) Among the contributions
that are accounted for by the expression, Eq. (\ref{MGTsu4.0}), it indicates
which one is retained by the OEM. ii) The energy difference, $%
E_{i}^{s}-E_{f}^{s}$, implies the single particle energies of nucleons in
the initial and final states. These ones involving the core particles, it
makes it clear that the various commutators appearing in Eq. (\ref{MGTsu4.0}%
) involve three and more body operators.

\section{THE EFFECTIVE NN POTENTIAL}

As mentioned in the introduction, we are motivated in this paper by two
different points. First, we want to study the contributions of the different
pieces of the nuclear force to the $\beta \beta 2\nu $-decay. Second, we
want to compare the two calculation methods presented in the previous
section. We will focus on the double beta decay of $^{48}Ca$ because it is a
nucleus which can be theoretically described in the nuclear shell model and
we can do reliable calculations with both methods. To accomplish our
objective, we use an analytical force which, therefore, could not be the
best one but, as we will observe later on, the results are good enough to
make it credible. In this way, we can easily connect and disconnect the
different pieces of the force and calculate the matrix elements of the
operators present in (\ref{MGToem1}). We have performed our calculations
using the OXBASH code (\cite{OXBASH}).

The shell model space is the full fp shell with the single particle energies 
$\epsilon _{f_{7/2}}=0,$ $\epsilon _{p_{3/2}}=2.1MeV,$ $\epsilon
_{p_{1/2}}=3.9MeV$ and $\epsilon _{f_{5/2}}=6.5MeV$. We have used for our
calculations the Bertsch-Hamamoto force \cite{Bertsch}. This force has a
central part which in momentum space is given by: 
\begin{align}
V_{c}\left( q\right) & =\left( \frac{f}{m_{\pi }}\right) ^{2}\left[ \frac{1}{%
3}\vec{\sigma}_{1}.\vec{\sigma}_{2}\,\vec{\tau}_{1}.\vec{\tau}_{2}\frac{%
m_{\pi }^{2}}{q^{2}+m_{\pi }^{2}}\right. +  \nonumber \\
& \left( a_{a}\Pi ^{S=0}\Pi ^{T=1}+b_{a}\Pi ^{S=1}\Pi ^{T=0}\right) \frac{%
m_{a}^{2}}{q^{2}+m_{a}^{2}}+  \nonumber \\
& \left( a_{b}\Pi ^{S=0}\Pi ^{T=1}+b_{b}\Pi ^{S=1}\Pi ^{T=0}\right) \left. 
\frac{m_{b}^{2}}{q^{2}+m_{b}^{2}}\right] ,  \label{B-H.Central}
\end{align}
with $f=0.97,$ $m_{\pi }$ the pion mass, $m_{a}=2.5%
%TCIMACRO{\unit{fm}}%
%BeginExpansion
\mathop{\rm fm}%
%EndExpansion
^{-1},\,m_{b}=4%
%TCIMACRO{\unit{fm}}%
%BeginExpansion
\mathop{\rm fm}%
%EndExpansion
^{-1},$ and a tensor part 
\begin{align}
V_{T}\left( q\right) & =-\left( \frac{f}{m_{\pi }}\right) ^{2}\frac{1}{3}%
\left( 3\vec{\sigma}_{1}.\vec{q}\,\vec{\sigma}_{2}.\vec{q}-\vec{\sigma}_{1}.%
\vec{\sigma}_{2}q^{2}\right) \hspace{2cm}  \nonumber \\
& \times \left[ \vec{\tau}_{1}.\vec{\tau}_{2}\left( \frac{1}{q^{2}+m_{\pi
}^{2}}-\frac{t_{1}}{q^{2}+m_{a}^{2}}\right) -\frac{t_{0}}{q^{2}+m_{a}^{2}}%
\right] \,.  \label{B-H.Tensor}
\end{align}
Bertsch and Hamamoto (B.H.) fitted the parameters of the force in order to
reproduce Reid soft-core G matrix elements and used this force to estimate
the Gamow-Teller strength at high excitation in $^{90}Zr$. The values for
these parameters are given in Table \ref{tabla1}. The main features of this
force are: (i) it contains the one-pion exchange, which governs the long
range part of the force, both in the central and the tensor terms; (ii) the
other terms of the central force are pure $S$ wave interaction; (iii) the
attraction in the $S=1,T=0$ channel is bigger than the one in the $S=0,T=1$
channel (thus the pairing in the $T=0$ channel will be greater than the
usual pairing in the $T=1$ channel); (iv) the B-H tensor force was fitted to
be used only for $L$ even waves (the $T=0$ channel) and in this order only
one of $t_{0,1}$ parameters is necessary. The authors of ref. \cite{Bertsch}
chose $t_{0}=0$ while the tensor interaction in the $T=1$ channel was
completely discarded. We have checked that this tensor force, as an
effective one, is consistent with the deuteron D-wave. In order to have a
simultaneous description of the three nuclei involved in the transition, $%
^{48}Ca$, $^{48}Sc$ and $^{48}Ti$, we also fitted the average Coulomb
displacement in Eq. (\ref{Coulomb Force}), $\Delta _{c}$, using the relative
position of the ground state of $^{48}Ca$ to that of $^{48}Ti$.

We nevertheless observe that the B-H force evidences some undesirable
features when it is applied to the study of the nucleus spectroscopy. For
instance, in the region of $^{48}Ca$ of interest here: (i) it gives a state
density at low energy larger than obtained with other standard potentials
like modified versions of the Kuo-Brown G-matrix interaction \cite{Brown} or 
\cite{Caurier1}; (ii) the splitting between the first $J=0^{+},\,T=1$ state
and the first $J=1^{+},\,T=0$ state of $^{42}Sc$ has the wrong sign; (iii)
if we extend the tensor force as it is in the original B-H force to the $T=1$
channel, it produces quite important matrix elements. What happens is that
this force must be used with some short range correlations which will
decrease its effective intensity. We will not introduce short range
correlations and for this reason and from the fact that a so simple force
cannot have unchangeable parameters in a large range of nuclei, we slightly
modified the original B-H parameters. (i) we fitted $a_{a}$ and $b_{a}$,
reproducing the relative position of the ground states of $^{48}Ca$, $%
^{48}Ti $ and the first $J=1^{+}$ state of $^{48}Sc.$ This change has
reduced the state density in $^{48}Sc$ and has also corrected the splitting
between the two first $0^{+}$ states of $^{42}Ca$ as well as the splitting
between the first $0^{+}$ and the first $1^{+}$ states of $^{42}Sc$ (these
states are important to determine the two-body effective interaction). (ii)
we changed $t_{0}$ and $t_{1}$ in such a way that the tensor matrix elements
between two particles in the fp shell coupled to $T=1$ has been strongly
reduced but without change in the matrix elements of particles coupled to $%
T=0.$ These parameters are also given in Table \ref{tabla1}\ as modified
B-H. Due to their limited number, our force is not the most realistic one.
Results presented here cannot therefore compete with other ones which rely
on a better force. As it can be seen from the results we obtain, they are
realistic enough however so that our schematic study makes sense and can
provide sensitive information.

In Table \ref{tabla2}, we present the first states with quantum numbers $%
1^{+}$ for $^{48}Sc.$ Notice that, with respect to the state density
argument presented above, this table is partly misleading. Higher energy
states should be included in the comparison.

We looked at the distribution of the Gamow-Teller strength for these forces
and compared it with a standard calculation performed with a modified
Kuo-Brown interaction \cite{Brown}. As it can be observed in Fig. \ref{fig2}%
, there is no difference between the strength calculated with the B-H or the
modified B-H interactions and that one using the potential of ref. \cite
{Brown}. The $\beta ^{+}$ strength from the final state has also been looked
at. It is shown in Fig. \ref{fig3} for the same models as mentioned above.
Its relevance has been mentioned several times in the literature and
re-emphasized recently in ref. \cite{Urin}. It represents an important
constraint. We observe that for the B-H and modified B-H potentials the
results are hardly distinguishable; for the modified K-B interaction more
strength is concentrated in the low energy region. The essential point is
that the $\beta ^{+}$ strength is large where the Gamow-Teller strength is
small and vice versa. Moreover, the contributions of the two regions could
be opposite in sign, which is not observed in the strength, making difficult
an accurate determination of the total matrix element.

We want to stress here two points. Our modification of the Bertsch-Hamamoto
force has not a fundamental origin. In that sense, we cannot say that this
modified force is better than the original one, but it gives a better
description of the spectra for the states of the nuclei we are considering.
On the other hand, the main motivation for using the Bertsch-Hamamoto force,
or a modified one, is that its analytical structure allows one to make a
simple analysis of the role of the different pieces while the total
transition matrix element is compatible with all previous calculations, as
it will be shown later.

\section{NUCLEAR POTENTIAL AND $\protect\beta\protect\beta2\protect\nu$-DECAY%
}

\begin{figure}
\begin{center}
\mbox{\epsfig{file=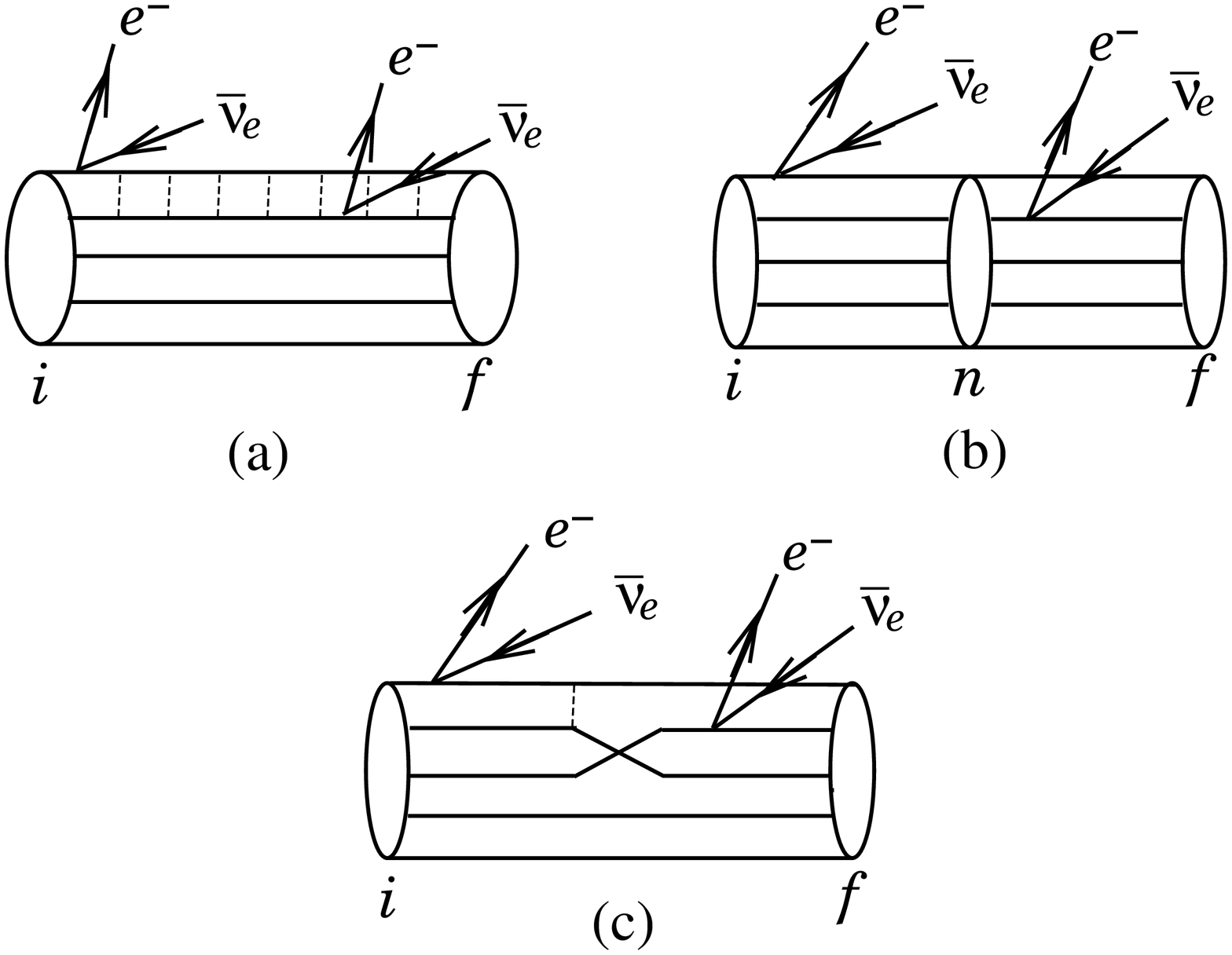,
width=0.7\linewidth,height=0.4\textheight}}
\end{center}
\caption{a) Diagrams included in the OEM approach. b) General diagram for
the Standard and $SU\left( 4\right) $ calculation. c) Example of exchange
diagram which is not included in the OEM.}
\label{fig1}
%\vspace{1cm}
%\centerline{\large P. Faccioli, M. Traini and V. Vento
%Phys. Lett. B
%}
\vspace{1cm}
\centerline{\bf \large FIGURE 1}

\newpage

\begin{center}
\mbox{\epsfig{file=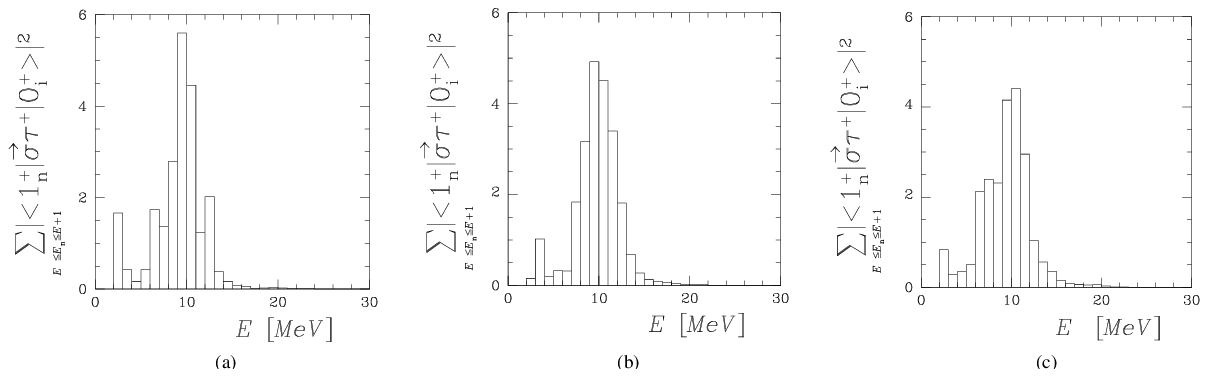,
width=1.0\linewidth,height=0.25\textheight}}
\end{center}
\caption{Calculated Gamow-Teller strength for the $^{48}Ca\rightarrow
^{48}Sc $ transition using the modified Kuo-Brown potential (case a), the
modified Bertsh-Hamamoto potential (case b), and the Bertsh-Hamamoto
potential (case c). Each histogram represents the total strength in 1 MeV
region versus the energy.}
\label{fig2}
%\vspace{1cm}
%\centerline{\large P. Faccioli, M. Traini and V. Vento
%Phys. Lett. B
%}
\vspace{1cm}
\centerline{\bf \large FIGURE 2}

\newpage

\protect
\begin{center}
\mbox{\epsfig{file=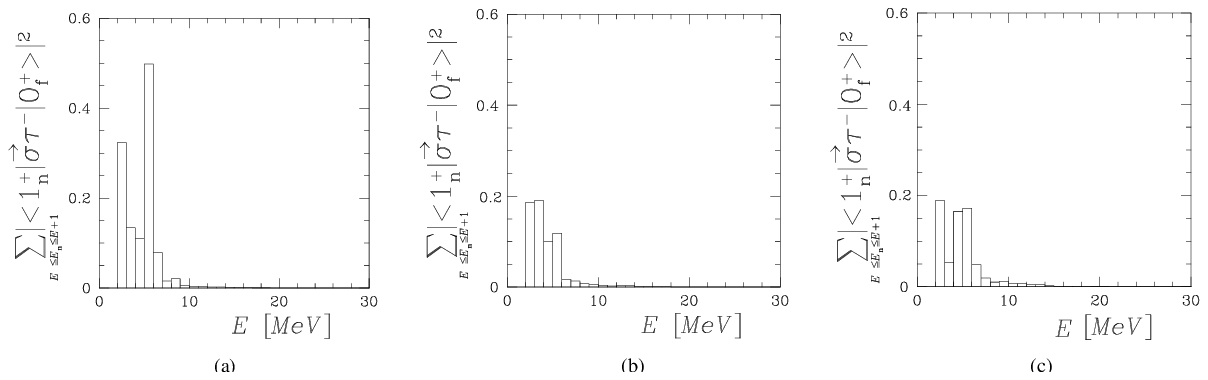,
width=1.0\linewidth,height=0.25\textheight}}
\end{center}
\caption{Calculated $\protect\beta ^{+}$ strength for the $%
^{48}Ti\rightarrow ^{48}Sc$ transition using the modified Kuo-Brown
potential (case a), the modified Bertsh-Hamamoto potential (case b), and the
Bertsh-Hamamoto potential (case c). Each histogram represents the total
strength in 1 MeV region versus the energy.}
\label{fig3}
\vspace{1cm}
%\centerline{\large P. Faccioli, M. Traini and V. Vento
%Phys. Lett. B
%}
\vspace{1cm}
\centerline{\bf \large FIGURE 3}

\newpage

\begin{center}
\mbox{\epsfig{file=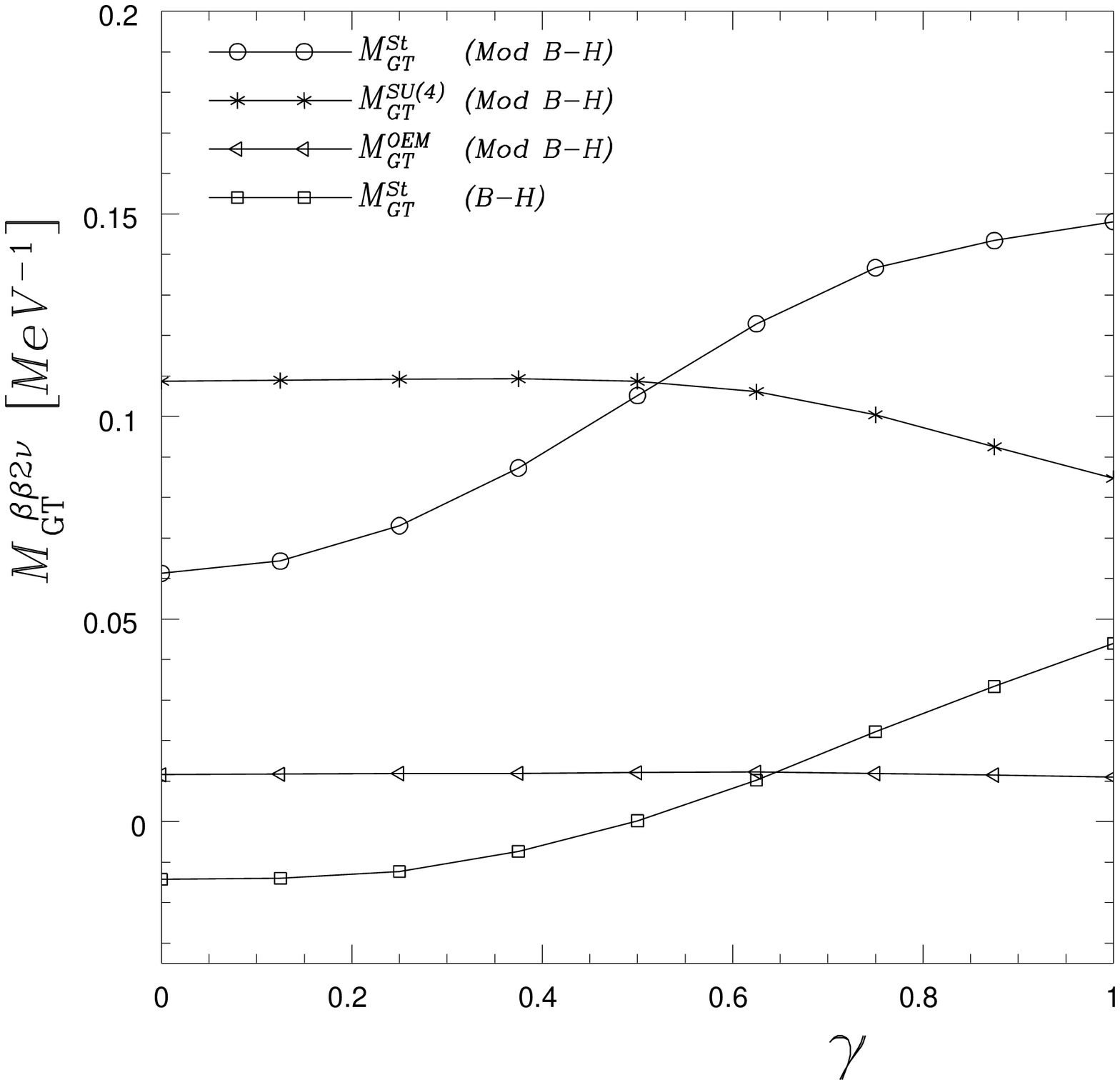,
width=0.6\linewidth,height=0.4\textheight}}
\end{center}
\caption{Double beta amplitude versus the spin-orbit potential. All curves
correspond to the modified B-H potential, except the $M_{{\cal GT}}^{St}$
with squares which corresponds to the B-H potential.}
\label{fig4}
\vspace{1cm}
%\centerline{\large P. Faccioli, M. Traini and V. Vento
%Phys. Lett. B
%}
\vspace{1cm}
\centerline{\bf \large FIGURE 4}

\newpage

\begin{center}
\mbox{\epsfig{file=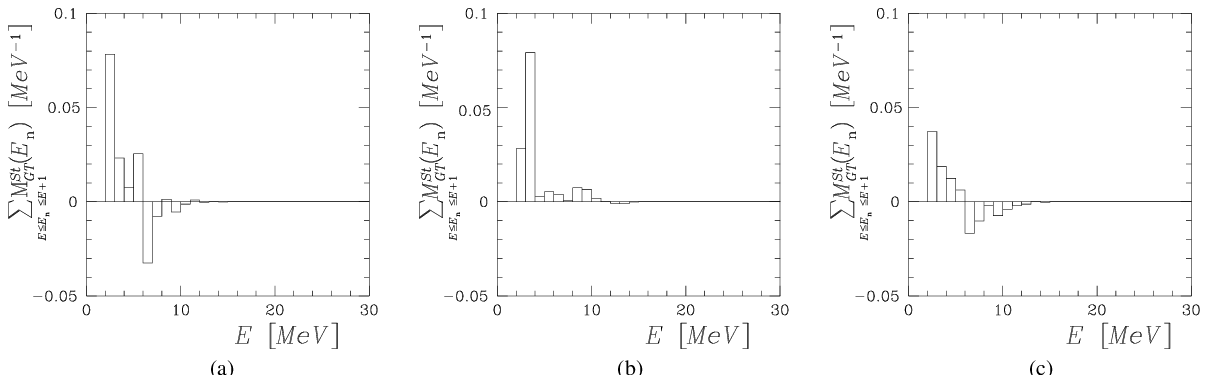,
width=1.0\linewidth,height=0.25\textheight}}
\end{center}
\caption{Detail of the contributions to the double beta amplitude, $M_{{\cal %
GT}}^{St}$, using the modified Kuo-Brown potential (case a), the modified
Bertsh-Hamamoto potential (case b), and the Bertsh-Hamamoto potential (case
c). Each histogram represents the total contribution in 1 MeV region versus
the energy.}
\label{fig5}
\vspace{1cm}
%\centerline{\large P. Faccioli, M. Traini and V. Vento
%Phys. Lett. B
%}
\vspace{1cm}
\centerline{\bf \large FIGURE 5}

\end{figure}

Our first study concerns the role of the central force in relation with the
contribution of the spin-orbit. In this order, we turned off the tensor
potential. The spin-orbit energies are multiplied by a factor $\gamma ,$
running from $0$ to $1,$ in such a way that, for $\gamma =0,$ only the
central potential contributes while, for $\gamma =1,$ the spin-orbit
splitting energies are completely accounted for. The results for the
different calculation methods (\ref{MGTst}), (\ref{MGToem}-\ref{MGToem2})
and (\ref{MGTsu4}) are given in Fig. \ref{fig4}.

Comparing $M_{{\cal GT}}^{St}$ for the B-H and modified B-H potentials in
absence of spin-orbit potential ($\gamma =0),$ we observe that small changes
in the values of $a_{a}$ and $b_{a}$\ make the double beta amplitude to go
through zero. Hence, the contribution of the central potential by itself is
not completely under control. This result points to the relative weight of
the forces in the $(S=1,T=0)$\ and $(S=0,T=1)$\ channels, which plays an
essential role in the present field.

Quite generally, interaction models based on nucleon-nucleon scattering data
have a strength in the $(S=1,T=0)$ channel bigger than in the $(S=0,T=1)$
one (as is the case of the B-H potential, see Table \ref{tabla3}), but most
effective nuclear potentials fitted to reproduce the spectra of nuclei give
a pairing for $(S=0,T=1)$ states stronger than for $(S=1,T=0)$ states.
Typically, the situation is characterized by nuclear matrix elements like
those displayed in Table \ref{tabla3}, calculated for the B-H and the
modified B-H potentials. It has a direct relationship to the relative weight
of the forces, $v_{\tau }\left( r\right) $\ and $v_{\sigma }\left( r\right) $%
, in Eq. (\ref{central pot}). The issue is an important one, which has a
close relationship to the sensitivity to the so-called $g_{pp}$ parameter
appearing in other approaches. As there, one has to hope that the fit of the
effective nuclear potential model to a few relevant experimental
informations will allow one to minimize uncertainties. In ref. \cite{Poves}%
,\ Poves et al. considered the same problem in terms of two factors, $%
\lambda _{01}$\ and $\lambda _{10}$\ , multiplying respectively the
strengths of the forces in the singlet and triplet spin channels. Starting
from a force that was already good, the variation for these factors is
actually smaller than what is suggested by the comparison of our matrix
elements given in Table \ref{tabla3} for the B. H. and the modified B. H.
forces. In ref. \cite{Schuck}, one can find a recent study on the $\left(
pn\right) $ pairing and the relevance of the point here underlined in heavy
nuclei, which are studied in the QRPA approach. This is also discussed in
ref. \cite{Engel-Dukelsky}. An argument is sometimes advocated for the
change of the relative strength of the forces in the singlet and triplet
spin channels when going from infinite nuclear matter to finite\ nuclei. It
relies on the effect of the spin-orbit force. The force in the singlet spin
channel is coupling preferentially particles with the same quantum numbers, $%
\left( j,l\right) $, whereas the force in the triplet spin channel rather
couples spin-orbit partners, the effective force is favored by the absence
of spin-orbit splitting in the first case while it is disfavored by its
presence in the other.

The role of the spin-orbit interaction has not received much attention in
the field, probably because it is known and is not considered as a free
parameter. In an approach based on the $SU(4)$ symmetry\ like that one
referred to here, it has some relevance since it is a piece of the
interaction that breaks the symmetry. Its importance can be seen by looking
at the dependence of $M_{{\cal GT}}^{St}$ on $\gamma $ shown in Fig. \ref
{fig4}. It represents a quite important contribution for both potentials. In
the B.H. case, it produces a change in sign while in the modified B. H.
case, it enhances the amplitude by a factor 2. Algebraically, the effect is
roughly the same and the difference by a factor 3 between the results for $%
\gamma =1$, $M_{{\cal GT}}^{St}\left( \text{B-H}\right) =0.044$ MeV$^{-1}$
and $M_{{\cal GT}}^{St}\left( \text{mod B-H}\right) =0.15$ MeV$^{-1},$ is
thus due to the central potential contribution, clearly over-estimated in
the modified B-H potential.

It is instructive to look at the detail of the contributions of the
intermediate states to the matrix element $M_{{\cal GT}}^{St}$, in Eq. (\ref
{MGTst}). This is given in Fig. \ref{fig5} for the various models whose
Gamow-Teller $\beta ^{-}$\ and $\beta ^{+}$\ strengths\ were shown in Figs. 
\ref{fig2} and \ref{fig3} respectively. For the B-H model (as well as the
K-B model), there are contributions with both signs, respectively located at
low and high energy. The dominant contribution in the low energy range is
indirectly an effect of the spin-orbit interaction which brings down some
states into this region and at the same time some strength. It is partly
cancelled by a contribution in the Gamow-Teller resonance region, which is
reminiscent of that one estimated in the $SU(4)$ symmetry approach (see
below) or that one calculated in \cite{Bernabeu}\ on the basis of the
dominance of this resonance in the sum entering Eq. (\ref{MGTst}). For the
modified B-H model, all contributions are positive. The low energy range one
has the same origin as above, whereas that one in the Gamow-Teller resonance
region has the opposite sign. This is due to the change in the relative
strengths of the forces in the singlet and triplet spin channels evidenced
by these models (Table \ref{tabla3}). As a result, the matrix element $M_{%
{\cal GT}}^{St}$\ for the modified B-H model is significantly larger.

We now focus on the modified B-H potential and compare $M_{{\cal GT}}^{St}$
with $M_{{\cal GT}}^{SU(4)}$. We observe that $M_{{\cal GT}}^{SU(4)}$ gives
a reasonable estimate of $M_{{\cal GT}}^{St},$ up to a factor 2. But this
nice result is partly due to the crossing of the two curves in Fig. \ref
{fig4}, $M_{{\cal GT}}^{St}$ and $M_{{\cal GT}}^{SU(4)},$ which makes their
difference to remain in a relatively small range. The accidental character
of the agreement is evidenced by looking at results for a different choice
of the central force. Thus, for the case of the B-H potential, $M_{{\cal GT}%
}^{SU(4)}$ varies from -0.072 MeV$^{-1}$ for $\gamma =0$ to -0.053 MeV$^{-1}$
for $\gamma =1$ while for the same potential $M_{{\cal GT}}^{St}$ takes
values from -0.014 MeV$^{-1}$ for $\gamma =0$ to 0.044 MeV$^{-1}$ for $%
\gamma =1.$ We must conclude that, even if $M_{{\cal GT}}^{SU(4)}$ is a good
estimate of the order of magnitude, it does not give the right sign and,
moreover, differences for the absolute value can be as big as a factor 2 or
more. The change in sign for $M_{{\cal GT}}^{SU(4)}$, when going from the
B-H to the modified B-H potential, is due to the relative value of the $T=1$
and $T=0$ pairing as is shown in Table \ref{tabla3}.

Looking at the dependence of $M_{{\cal GT}}^{SU(4)}$ on $\gamma ,$ we
observe that its value\ is relatively stable. When $\gamma $ runs from 0 to
1, the spin-orbit contribution to the wave function is included to all
orders in $M_{{\cal GT}}^{SU(4)}$ but its vertex contribution, through the
operator in (\ref{MGT}), is not considered. What shows the evolution of $M_{%
{\cal GT}}^{SU(4)}$ is that this vertex contribution is the dominant one.
Naively, we could conclude that a better estimate is to approach the
Hamiltonian in (\ref{MGT}) by $H=H_{so}+\Delta _{c}T_{3},$ but this
contribution vanishes. In fact, the result we obtained for $M_{{\cal GT}%
}^{St}$\ implies a large interference between the spin-orbit potential and
the central potential. This is perhaps a consequence of the strengthening of
the force in the singlet spin channel with respect to the triplet one, which
we mentioned above as being indirectly due to the spin-orbit force.

Looking now at $M_{{\cal GT}}^{OEM},$ we must conclude that its value is
mostly independent of the spin-orbit potential. This statement is also true
for the B-H potential. In that case, $M_{{\cal GT}}^{OEM}$ runs from -0.011
MeV$^{-1}\text{\ }$for $\gamma =0$ to -0.010\ MeV$^{-1}\text{\ }$for $\gamma
=1.$ The difference in sign for $M_{{\cal GT}}^{OEM}$ calculated with the
B-H and the modified B-H potential is again due to the relative value of the 
$T=1$ and $T=0$ pairing. In this case, $M_{{\cal GT}}^{OEM}$ is not a good
estimate; neither the sign nor the absolute value are well reproduced. In
respect to the $M_{{\cal GT}}^{OEM}$ calculation, we must emphasize that the
Coulomb force cannot be neglected. If we use (\ref{MGToem1}) instead of (\ref
{MGToem2}) our results for $M_{{\cal GT}}^{OEM}$ vary from 0.0005\ MeV$^{-1}%
\text{\ }$for $\gamma =0$ to 0.0003\ MeV$^{-1}\text{\ }$for $\gamma =1,$
evidencing an absolutely non sense result. We observe that $M_{{\cal GT}%
}^{OEM}$ and $M_{{\cal GT}}^{SU\left( 4\right) }$ have the same sign, in
apparent contradiction with what was said in section 3. What happens is that 
$M_{{\cal GT}}^{OEM}$ has a peculiar behavior when the $SU\left( 4\right) $
breaking part of the central force is reduced, crossing the zero and
changing sign when we multiply (\ref{H1}) by a factor $\kappa $ and study
the limit of $\kappa $ going to zero.

Beside the role of the spin-orbit potential, whose importance has been
discussed above, we also considered the contribution of the tensor
potential. We found that this one does not change the $\beta \beta 2\nu $%
-transition amplitude in a significant way. Only a slight decrease was
observed. This can be seen as due to an effective decrease of the spin-orbit
interaction which is in fact observed around $^{48}Ca$\ and has been
attributed to the tensor force in the past \cite{Bouyssy}. Our full results
so obtained are: 
\begin{equation}
M_{{\cal GT}}^{St}=0.135\text{MeV}^{-1,}
\end{equation}
\begin{equation}
M_{{\cal GT}}^{SU(4)}=0.084\text{MeV}^{-1},
\end{equation}
\begin{equation}
M_{{\cal GT}}^{OEM}=0.011\text{MeV}^{-1},
\end{equation}
for the modified B-H potential and 
\begin{equation}
M_{{\cal GT}}^{St}=0.032\text{MeV}^{-1},
\end{equation}
\begin{equation}
M_{{\cal GT}}^{SU(4)}=-0.051\text{MeV}^{-1},
\end{equation}
\begin{equation}
M_{{\cal GT}}^{OEM}=-0.008\text{MeV}^{-1},
\end{equation}
for the B-H potential.

Previous results for the amplitude calculated in the standard way are
summarized in Table \ref{tabla4}. They can be compared to the experimental
value $M_{{\cal GT}}=0.074%
%TCIMACRO{\QATOPD. . {+0.012}{-0.015}}%
%BeginExpansion
{+0.012 \atopwithdelims.. -0.015}%
%EndExpansion
%TCIMACRO{\unit{MeV}}%
%BeginExpansion
\mathop{\rm MeV}%
%EndExpansion
^{-1}$, also given in the Table. As it can be seen, our results are between
a factor 2 too high for the modified B-H potential and a factor 2 too small
for the original B-H potential. In view of the simplicity of the force used
in present investigations, which has allowed us to study the role of its
different pieces, results can be considered as reasonable. Summarizing the
main features, it can be noticed that the amplitudes, $M_{{\cal GT}}^{St}$
and $M_{{\cal GT}}^{SU\left( 4\right) }$, are shifted upwards by roughly the
same amount when going from the B-H to the modified B-H potential. This is
related to the change in the relative strengths of the force in the $S=0,T=1$
and $S=1,T=0$ channels. The difference between these two amplitudes,
indirectly due to the spin-orbit force, is relatively insensitive to this
modification. Both effects are important to get a value that compares to the
experimental one. The results for $M_{{\cal GT}}^{OEM}$ also show some
sensitivity but are out of range in any case. Results of previous
calculations employing other methods are given in Table V. It is seen that
the above values fall in the range of the more realistic estimates,
especially that one by Caurier et al. \cite{Caurier2}, $M_{{\cal GT}%
}^{St}=0.065%
%TCIMACRO{\unit{MeV}}%
%BeginExpansion
\mathop{\rm MeV}%
%EndExpansion
^{-1}$, which is probably the most elaborate one. This indicates that our
study, though schematic, deals with the real problems underlying the
calculation of the $M_{{\cal GT}}$ amplitude.

\section{CONCLUSIONS}

We considered the $\beta \beta 2\nu $-decay process in the nucleus of $%
^{48}Ca$ with the aim to analyze different methods used in estimating the
corresponding transition amplitude or to study the role of different
components in the nuclear interaction. In this sense, $^{48}Ca$ is used as a
theoretical laboratory for testing various approaches. In all cases, we have
performed our calculations in the full $f-p$ shell. In order to discuss the
different terms of the potential, we used the B-H potential, which is
analytical, and adapted it to our nuclei giving rise to what we called
modified B-H potential. Our conclusions do not depend on particular aspects
of one or the other potential.

First, we confirmed the strong sensitivity of the $\beta \beta 2\nu $\
amplitude to the relative strength of the potentials in channels $\left(
S=1,T=0\right) $ and $\left( S=0,T=1\right) .$ This is well a known result
and its importance has been emphasized in the QRPA calculations (in a
different language, this was firstly pointed out by \cite{Vogel} as well as 
\cite{Krmpotic}). But our main conclusion is that the central force alone,
due to this cancellation, does not provide the leading contribution to be
considered in the calculation of the $\beta \beta 2\nu $-transition
amplitude.

From our results, the single particle spin-orbit force appears to be the
main ingredient in determining the actual value of the amplitude.
Nevertheless, this term of the potential alone is not sufficient as the
total amplitude then vanishes. Any sensitive estimate of the amplitude
requires the interference between this spin-orbit term and the two-body
parts of the strong potential. In this interference, operators involving
three, or even more, nucleons could appear.

Concerning the tensor potential, we did not find it was relevant for the $%
\beta \beta 2\nu $-decay amplitude.

The results obtained in the OEM approximation are far from the exact
calculation. The OEM is not an approach under control, as already mentioned
in the literature. Sizeable corrections come from three or more body
operators but the non-convergent character of the expansion don't let much
hope that the corrections are manageable. We have put in evidence that the
OEM has a wrong $SU(4)$ limit and this difference is also originated from
the many body operators. Moreover, we observed that the Coulomb potential
cannot be neglected at all in this scheme.

For the future, one can imagine to improve the approach based on the $SU(4)$%
\ symmetry. It is not clear however whether accounting for the spin-orbit
splitting is feasible while keeping a rather simple form for the expression
of the transition amplitude. Another issue concerns the sign of the
contributions of the different intermediate states to the total $\beta \beta
2\nu $-transition amplitude. The $\beta ^{-}$\ and $\beta ^{+}$ Gamow-Teller
excitation\ from the initial and the final states respectively only know
about the magnitude. The interesting question is to know whether this
information together with the knowledge about the total transition
amplitude, $M_{{\cal GT}}^{St}$, can provide a clue as to the constructive
or destructive character of the partial contributions, as exemplified by two
estimates presented in this work.

This work has been partially supported by DGESIC (Spain) under contract N%
%TCIMACRO{\UNICODE{0xba} }%
%BeginExpansion
${{}^o}$%
%EndExpansion
PB97-1401-C02-01.

\begin{table}[tbp]
\begin{tabular}{|c|ccccccc|}
\hline
& $a_{a}$ & $b_{a}$ & $a_{b}$ & $b_{b}$ & $t_{1}$ & $t_{0}$ & $\Delta
_{c}(MeV)$ \\ \hline
Bertsch-Hamamoto & -8.28 & -14.33 & 6.56 & 11.20 & 0.89 & 0 & 5.352 \\ \hline
Modified B-H & -8.89 & -14.13 & 6.56 & 11.20 & 1.335 & 1.335 & 5.512 \\ 
\hline
\end{tabular}
\caption{Parameters entering the nuclear force, equations (\ref{B-H.Central}%
, \ref{B-H.Tensor}).}
\label{tabla1}
\end{table}

\begin{table}[tbp]
\begin{tabular}{l||l||lllllll}
\hline
Exp. & 6.51 & 0.46 & 0.54 & 0.64 & 0.74 & 1.19 & 1.48 & 1.66 \\ \hline
B-H & 4.96 & 0.27 &  & 0.68 & 0.83 & 1.17 & 1.44 & 1.67 \\ \hline
Modified B-H & 6.50 &  & 0.55 &  & 0.93 & 1.28 &  & 1.60 \\ \hline
Modified K-B ref \cite{Brown} & 6.49 &  &  & 0.63 &  & 0.97 & 1.35 & 1.67 \\ 
\hline
Modified K-B ref \cite{Caurier1} & 6.38 &  & 0.52 &  &  & 1.07 & 1.56 &  \\ 
\hline
\end{tabular}
\caption{In the first column, we give the position of the first state in $%
^{48}Sc$ relatively to the fundamental state in $^{48}Ti.$ The following
columns contain the excited energies of the following $1^{+}$ states in $%
^{48}Sc$ relatively to the first $1^{+}$ state.}
\label{tabla2}
\end{table}

\begin{table}[tbp]
\begin{tabular}{|c|cc|}
\hline
& $S=0\,T=1$ & $S=1\,T=0$ \\ \hline
B-H & -2.98 & -4.87 \\ 
Modified B-H & -5.03 & -4.24 \\ \hline
\end{tabular}
\caption{Diagonal matrix elements $(f^2 L=0, ST)$ for the cases $(S=0,T=1)$
and $(S=1,T=0)$ for the central part of the B-H and the modified B-H
potentials, in MeV. The relative weight of this type of matrix elements
determines the sign of the corresponding part of the force to the double
beta amplitude}
\label{tabla3}
\end{table}

\begin{table}[tbp]
\begin{tabular}{|l|l||l|l||l|l|}
\hline
& $M_{{\cal GT}}^{St}($MeV$^{-1})$ &  & $M_{{\cal GT}}^{St}($MeV$^{-1}) $ & 
& $M_{{\cal GT}}^{St}($MeV$^{-1})$ \\ \hline
Mod. B-H & 0.135 & ref \cite{Caurier2} & 0.065 & ref \cite{Ogawa} & 0.089 \\ 
\hline
B-H & 0.032 & ref \cite{Engel1} & 0.043 & ref \cite{Radha} & 0.15$\pm 0.07$
\\ \hline
Experimental & 0.074$%
%TCIMACRO{\QATOPD. . {+0.012}{-0.015}}%
%BeginExpansion
{+0.012 \atopwithdelims.. -0.015}%
%EndExpansion
$ & ref \cite{Brown} & 0.055 &  &  \\ \hline
\end{tabular}
\caption{Double beta amplitudes for the two $\protect\nu$ mode. The second
row of the first column corresponds to the modified Bertsch-Hamamoto force,
the third row of the first column to the original Bertsch-Hamamoto and the
fourth row of the first column to the experimental result. All the other
columns correspond to usual shell model calculations except for ref. 
\protect\cite{Radha} which corresponds to a shell model Monte Carlo
technique. The experimental result corresponds to \protect\cite{Balysh}. We
have taken $g_{A}=1.0$ }
\label{tabla4}
\end{table}

\begin{table}[tbp]
\begin{tabular}{|l|l|l|l|l|l|}
\hline
& Mod. B-H & B-H & ref \cite{Engel1} & ref \cite{Ching} & ref\cite{Noguera}
\\ \hline
$M_{{\cal GT}}^{SU(4)}$ & 0.084 & -0.051 &  &  & -0.062 \\ \hline
$M_{{\cal GT}}^{OEM}$ & 0.011 & -0.008 & -0.012 & -0.020/-0.035 &  \\ \hline
\end{tabular}
\caption{Results for $M_{{\cal GT}}^{SU(4)}$ and $M_{{\cal GT}}^{OEM}$ in $%
MeV^{-1}$}
\label{tabla5}
\end{table}

\newpage

\end{document}